\documentclass[useAMS]{mn2e}
\usepackage{lscape}
\usepackage{rotating}
\usepackage{amssymb}

\def\av{\hbox{A$_{\rm V}$}}

\def\msun{\hbox{M$_\odot$}}
\def\zsun{\hbox{Z$_\odot$}}

\def\t4{\hbox{t$_{\rm 4}$}}

\def\cm3{\hbox{cm$^{-3}$}}
\input psfig.sty
\input epsf.sty
\usepackage{graphicx}
\usepackage{epsfig}
\title[Using YMCs to Test the FRMS Scenario]
{Constraining Globular Cluster Formation Through Studies of Young Massive Clusters - IV. Testing the Fast Rotating Massive Star Scenario}
\author[Bastian et al.] {N. Bastian$^{1}$, K. Hollyhead$^{1}$ \& I. Cabrera-Ziri$^{1}$\\
$^{1}$ Astrophysics Research Institute, Liverpool John Moores University, 146 Brownlow Hill, Liverpool L3 5RF, UK\\
}
\date{Accepted. Received; in original form}
\pagerange{\pageref{firstpage}--\pageref{lastpage}}
\pubyear{2014}
\begin{document}
\maketitle
\label{firstpage}
\begin{abstract}
One of the leading models for the formation of multiple stellar populations within globular clusters is the ``Fast Rotating Massive Star" (FRMS) scenario, where the ejecta of rapidly rotating massive stars is mixed with primordial material left over from the star-formation process, to form a second generation of stars within the decretion discs of the high mass stars.  A requirement of this model, at least in its current form, is that young massive (i.e. proto-globular) clusters are not able to eject the unused gas and dust from the star-formation process from the cluster for $20-30$~Myr after the formation of the first generation of stars, i.e. the cluster remains embedded within the gas cloud in which it forms.  Here, we test this prediction by performing a literature search for young massive clusters in nearby galaxies, which have ages less than $20$~MyrΤ that are not embedded.  We report that a number of such clusters exist, with masses near, or significantly above $10^6$~\msun, with ages between a few Myr and $\sim15$~Myr, suggesting that even high mass clusters are able to clear any natal gas within them within a few Myr after formation.  Additionally, one cluster, Cluster 23 in ESO~338-IG04, has a metallicity below that of some Galactic globular clusters that have been found to host multiple stellar populations, mitigating any potential effect of differences in metallicity in the comparison.  The clusters reported here are in contradiction to the expectations of the FRMS scenario, at least in its current form.  %The observations presented here also impact the ``AGB scenario", as any pristine gas within or near the proto-globular cluster is removed to distances of $>200$~pc, where it is likely unbound from the cluster, making it difficult for the cluster to re-accrete the gas, which is necessary in order to match the observed abundance patterns in present day globulars.%bservations of a $\sim10^6$~\msun\ cluster (stellar mass) with an age of $1-2$~Myr with near-zero reddening (i.e. it has already emerged from its progenitor cloud).  Such a cluster is in contradiction with the ``spin-star" scenario.

\end{abstract}

\section{Introduction}
\label{sec:intro}

In this series of papers, we are using the properties of young massive clusters (YMCs; a.k.a., young globular clusters) to place constraints on scenarios for the formation of globular clusters (GCs), in particular, the origin of the observed multiple stellar populations within GCs.  In the first paper of the series (Bastian et al.~2013a, hereafter Paper~I), we analysed the integrated optical spectra or resolved stellar photometry of 129 Galactic and extragalactic clusters with masses between $10^4$ and $10^8$~\msun\ and ages between $10$ and $1000$~Myr to search for evidence of ongoing star-formation within them (as predicted by various globular cluster formation scenarios, e.g., D'Ercole et al.~2008; Goudfrooij et al.~2011).  No clusters with ongoing star-formation were found, down to limits of 1-2\% of the cluster mass, strongly constraining scenarios that invoke extended periods of star-formation within clusters.  In the second paper (Cabrera-Ziri et al.~2014, hereafter Paper~II), we used the integrated spectrum of Cluster 1 in NGC~34, a $\sim100$~Myr cluster with a mass of $\sim2\times10^7$\msun\ to estimate its star-formation history, to search for evidence of multiple discreet bursts, predicted by some theories of cluster formation.  The cluster was found to be well approximated by a single stellar population, with no evidence of a secondary burst down to mass ratios of $10-20$\% of the cluster mass.  In the third paper (Bastian \& Strader~2014, hereafter Paper~III), we searched for reservoirs of gas/dust within YMCs in the LMC (with masses between $10^4 - 10^5$\msun, and ages of $\sim15-300$~Myr) , which were predicted to exist in order to form secondary populations of stars within clusters (e.g., Conroy \& Spergel~2011).  No clusters with significant amounts of gas within were found, with observational limits of $M_{\rm H{\sc I}} \ll 0.1 \times M_{\rm stellar}$, in tension with previous predictions.

None of the models put forward to explain the multiple populations in GCs explicitly invoke special conditions, i.e. conditions only found in the early universe, suggesting that the same mechanisms should be operating in young massive clusters today.  This makes YMCs ideal places to test GC formation theories (e.g., Sollima et al.~2013).  Additionally, since both metal rich and metal poor GCs have been observed to host multiple stellar populations, which were likely formed in very different environments and at different redshifts (e.g., Brodie \& Strader~2006; Kruijssen~2014), it appears likely that the process of the formation of multiple stellar populations is related to the clusters themselves, and not their host environment (cf.~Renzini~2013).

In the current study, we turn our attention to another of the leading models for the formation of multiple stellar populations within globular clusters, namely the ``Fast Rotating Massive Star" scenario (FRMS).  In this model, the ejecta of rapidly rotating massive stars is mixed with primordial material left over from the star-formation process, to form a second generation of stars within the decretion discs of the high mass stars (Prantzos \& Charbonnel~2006; Decressin et al.~2007a,b).  This scenario has recently been expanded by Krause et al.~(2012; 2013),  in an attempt to develop a full and coherent model for the formation and early evolution of GCs.  It is this expanded model that we aim to test in the current work.

In this model, GCs initially form a mass-segregated, single generation of stars (the first generation), with a given star-formation efficiency.  The gas not used in the formation of the first generation of stars cannot be expelled from the cluster, due to the deep gravitational potential, although holes in the gas are made surrounding each of the massive stars.  However, dense, optically thick material surrounds each of the holes, and in the outer regions of the cluster (i.e., outside a half-light radius, where there are no massive stars) the clusters remain embedded in their natal gas.

The cluster remains in this embedded state for $\sim30$~Myr, when accretion onto dark remnants provides enough energy to unbind and remove the gas from the cluster.  Hence, a clear prediction of the FRMS scenario (at least this version of it) is that massive clusters should remain embedded for the first $\sim30$~Myr of their lives.  For lower mass clusters ($M_{\rm cluster} \lesssim 10^5$~\msun), the embedded phase lasts for much less, just a few Myr (e.g., Seale et al.~2012; Longmore et al.~2014; Hollyhead et al.~in prep.).  However, it is unclear if this also holds for more massive clusters, which are more akin to the young globular clusters ($M > 10^6$~\msun).  Theoretical estimates, based on the notion that radiation pressure is the dominant feedback mechanism in YMCs (e.g., Murray et al.~2010), suggest that the gas removal timescale in dense clusters should be largely independent of the cluster mass (e.g., Kruijssen~2012).  Here we explicitly test this prediction observationally.

In the previous papers in the series, particularly papers I and II, we tested the predictions of models for the formation of multiple populations in GCs that invoked the formation of a second (or further) generation after $10-30$~Myr.  No evidence for such secondary bursts were found, leading us to conclude that models that invoke a rapid secondary burst ($<10$ Myr - e.g., K13) or those that invoke only a single star-formation episode (e.g., the early disc accretion scenario - Bastian et al. 2013b) were favoured.  Recent work on the stellar populations within dwarf galaxies have provided complimentary results.  Larsen et al. (2012, 2014) studied the metallicity distributions of field stars and globular clusters in three dwarf galaxies.  The authors found that below a metallicity of [Fe/H]$\sim2$, GCs make up a significant amount of the total stellar mass  ($20-50$\%) of the galaxy.  This places strict constraints on how much more massive GCs could have been at birth, relative to their present mass, and requires extreme stellar initial mass function variations in the second generation (only stars with masses less than $0.8$~\msun\ are able to form) in order to match the observations (e.g., Prantzos \& Charbonnel~2006; D'Antona et al.~2013).  These observational constraints appear to be in conflict with predictions of models such as the AGB and FRMS scenarios, which require clusters to have been $5-10$ times more massive than their current mass.  The early disc accretion scenario does not require clusters to have been more massive in the past, hence is not in conflict with the above observations, although many caveats remain to be tested for that scenario (Bastian et al.~2013b).

In this paper we search for young, high mass clusters that are not embedded in order to quantitatively test the FRMS scenario as put forward in Krause et al. (2012, 2013).  We primarily use data from the literature, although we supplement this with archival imaging when required.  The models of Krause et al.~(2013) adopt a stellar mass of the cluster of $3\times10^6$\msun, so we will focus our attention on clusters near or above that mass.  In order to determine if a cluster is still embedded, we will use its estimated extinction ($A_V$) as well as search for evidence that the cluster has had a significant affect on the surrounding ISM. 

The paper is organised as follows.  In \S~\ref{sec:lit} we present a number of studies from the literature of the properties of YMCs in nearby dwarf and spiral galaxies.  In \S~\ref{sec:t352} we study two clusters in the Antennae merging galaxies in more detail and look at their effect on the surrounding ISM.  A particularly high mass and young cluster in the dwarf starburst galaxy ESO~338-IG04 is studied in detail in \S~\ref{sec:cluster23}, and in \S~\ref{sec:conclusions} we present a discussion of our results as well as our conclusions.
%\section{Observations and Techniques}
%\label{sec:obs}

\section{Clusters from the literature}
\label{sec:lit}

We begin by looking at a sample of well studied high mass clusters in nearby dwarf and spiral galaxies.  One such cluster resides in a large cluster complex in the spiral galaxy, NGC~6946.  The cluster has been studied extensively with HST-based photometry along with ground based high-resolution spectroscopy.  The cluster has an age of $\sim12$~Myr and a dynamically determined mass of $1.7(\pm0.5)\times10^6$\msun\ (e.g., Larsen, Brodie, \& Hunter~2004 - potentially higher as the luminosity profile was artificially truncated at $65$~pc due to the shallow profile index).  The cluster is surrounded by a shell of material, seen in dust and H$\alpha$, with a radius of 300~pc (Elmegreen, Efremov \& Larsen~2000).  Comparison of the integrated photometry of the cluster with simple stellar population (SSP) models of the appropriate age leads to an extinction estimate of $A_B = 0$ (Larsen et al.~2006a).  The metallicity of the cluster is approximately 0.5~\zsun\ (Larsen et al.~2006b; Gazak et al.~2014).

%NGC~6946-1447 - Larsen et al.~(2004), 15 Myr, $1.7\times10^6$~\msun\ (potentially higher as the profile was artificially truncated at $65$~pc), solar metallicty (Gazak et al. 2014), extinction $A_B = 1.3$.  Surrounded by a bubble (seen in dust and H$\alpha$ emission) with a radius of 300~pc - Elmegreen, Efremov, \& Larsen~2000

NGC~1569 is a nearby dwarf starburst galaxy that hosts a number of young high mass clusters.  Emission line studies have found an ISM abundance of $(12 + log(O/H)=8.19)$ (Kobulnicky \& Skillman~1997), or $\sim0.4$~\zsun.  Two clusters near the galactic centre have been studied in detail, cluster A with an age of $5-7$~Myr (Maoz et al.~2001) and a dynamical mass of $1.3 (\pm0.2)\times10^6$\msun\ (Smith \& Gallagher~2001) and cluster B with an age of 16~Myr and a mass of $1.2\times10^6$\msun\ (Larsen et al.~2008; 2011).  Both clusters are exposed (i.e., they have emerged from their natal cloud) and have have blown large holes in the ISM (Hunter et al.~2000), potentially truncating the starburst episode (Bastian~2008).

NGC~1705-1 is a massive cluster residing near the centre of the nearby dwarf galaxy, NGC~1705.  The cluster has little or no extinction (showing that it is clearly exposed), an age of 10-15~Myr, and a mass of $1.1\times10^6$~\msun (Heckman \& Leitherer~1997; Larsen et al.~2011).  The FWHM of the cluster is only 0.9~pc, making it one of the densest clusters studied to date (Larsen et al.~2011).  Due to the youth of the cluster, its metallicity is expected to be similar to that of H{\sc ii} regions within the galaxy, which have estimated abundances of $\sim1/3$~Z$_{\odot}$.  Not only has this cluster cleared out any gas within it, and in the nearby vicinity, the cluster is thought to be playing a dominant role in the expanding bi-polar supergalactic wind emanating from the galaxy (Meurer et al. 1992; Heckman \& Leitherer~1997).

Moll et al.~(2007) have studied a young massive cluster, Cluster 1, in NGC~1140, a starburst galaxy with LMC metallicity at a distance of $20$~Mpc.  The cluster has an age of $5\pm1$~Myr as determined through comparison of the observed photometry with SSP models.  Additionally, the cluster displays strong Wolf-Rayet features in the integrated spectrum, confirming the photometric age.  The photometrically determined mass of the cluster is $1.1\pm0.3 \times10^6$\msun, while the dynamically determined mass is $10\pm3 \times 10^6$\msun\ (Moll et al.~2007).  This difference is likely due to the influence of binaries on the observed velocity dispersion of young clusters (Gieles et al.~2010).  Even at this young age and high mass, the cluster is clearly exposed ($A_V < 0.6$~mag) and has cleared out much of the surrounding ISM (Moll et al.~2007).

\begin{table*}
\caption{The young massive clusters discussed in the current work.  The references are: 1) Larsen et al.~2001; 2) Larsen et al.~2004; 3) Larsen et al.~2006a; 4) Larsen et al.~2006b; 5) Gazak et al.~(2014); 6) Kobulnicky \& Skillman~1997; 7) Smith \& Gallagher~2001; 8) Maoz et al.~2001; 9) Larsen et al.~2008; 10) Larsen et al.~2011; 11) Heckman \& Leitherer~1997; 12)  Moll et al.~2007; 13) Whitmore et al.~(1999); 14) Whitmore et al.~(2010); 15) Bastian et al.~(2009); 16) Mengel et al.~(2008); 17) Trancho et al.~(2007); 18) {\"O}stlin et al.~(2003); 19) {\"O}stlin et al.~(2007). $^{a}$These values are the FWHM of the clusters, as effective radii are highly uncertain due to their shallow luminosity profiles.  $^{b}$This estimate is uncertain due to the uncertainty in the effective radius measurement, caused by the shallow luminosity profile.  $^{c}$See text for a discussion of the clusters' impact on the surrounding ISM.}
\label{tab:clusters}
\begin{tabular}{lcccccccc}
\noalign{\smallskip}
\hline
\hline
\noalign{\smallskip}
Cluster &  Galaxy & Age & Mass  & Mass & R$_{\rm eff}$ & Radius of Hole/shell & Metallicity & Reference\\
 & & [Myr]  & [10$^5$~\msun] & [10$^5$~\msun] & [pc] & [pc] \\
 & & & dynamical & photometric & & &\\
\hline
\noalign{\smallskip}
1447 & NGC~6946 &$12.5^{+2.5}_{-2.5}$& $17^{+5}_{-5}$ & $8^{+4}_{-4}$ & $10.2^{+1.6}_{-1.6}$& 300& $0.5~Z_{\odot}$& 1,2,3,4,5\\
A & NGC~1569 &$6^{+1}_{-1}$&$13^{+2}_{-2}$&$7.6^{+3.8}_{-3.8}$ &0.9$^{a}$& -$^{c}$&$0.4~Z_{\odot}$ & 6,7,8,10\\
B & NGC~1569 &$15^{+5}_{-5}$& $6.8^{+1.1}_{-1.1}$& $14^{+7}_{-7}$&1.4$^{a}$&-$^{c}$&$0.4~Z_{\odot}$ & 6,9,10\\
1 & NGC~1705 &$12.5^{+2.5}_{-2.5}$ & $4.8^{+1.2}_{-1.2}$$^{b}$& $11^{+0.5}_{-0.5}$& 0.9$^{a}$& -$^{c}$&$0.33~Z_{\odot}$& 7,10,11\\
1 & NGC~1140 & $5^{+1}_{-1}$& $100^{+30}_{-30}$ & $11^{+3}_{-3}$& $8^{+2}_{-2}$&-$^{c}$ &$0.5~Z_{\odot}$& 12\\
T352/W38220 & The Antennae & $4^{+2}_{-2}$ & - & 9.2$^{+4.6}_{-4.6}$  & 2.4$^{+1.5}_{-1.0}$ & 80 & $Z_{\odot}$ & 13,14,15  \\
Knot S & The Antennae & $5^{+1}_{-1}$ & 30$^{+12}_{-10}$ & 16$^{+8}_{-8}$& $8.0^{+1.5}_{-1.5}$ & 200 &  $Z_{\odot}$ & 13,16 \\
T2005 & NGC~3256 & $<5$ & - & 14$^{+7}_{-7}$  & -&  40 & $1.3~Z_{\odot}$ & 17 \\
Cluster 23 & ESO~338-IG04 & $6^{+4}_{-2}$& 130$^{+30}_{-30}$ & 50$^{+25}_{-25}$  & $5.2^{+1.0}_{-1.0}$ & 120-200&$0.2 Z_{\odot}$ &  18,19\\ 

%$J$ & 20$\times$60s  & 0\myarcsec36$-$0\myarcsec67\\
%$H$ & 60$\times$60s  & 0\myarcsec33$-$0\myarcsec68\\
%$K$ & 48$\times$60s  & 0\myarcsec37$-$0\myarcsec72\\
%$BrG$ & 16$\times$300s & 0\myarcsec37$-$0\myarcsec59\\
\hline
\end{tabular}
\end{table*}

\section{Clusters in Galaxy Mergers}
\label{sec:t352}

\subsection{The Antennae}

\begin{figure}
\centering
\includegraphics[width=6cm]{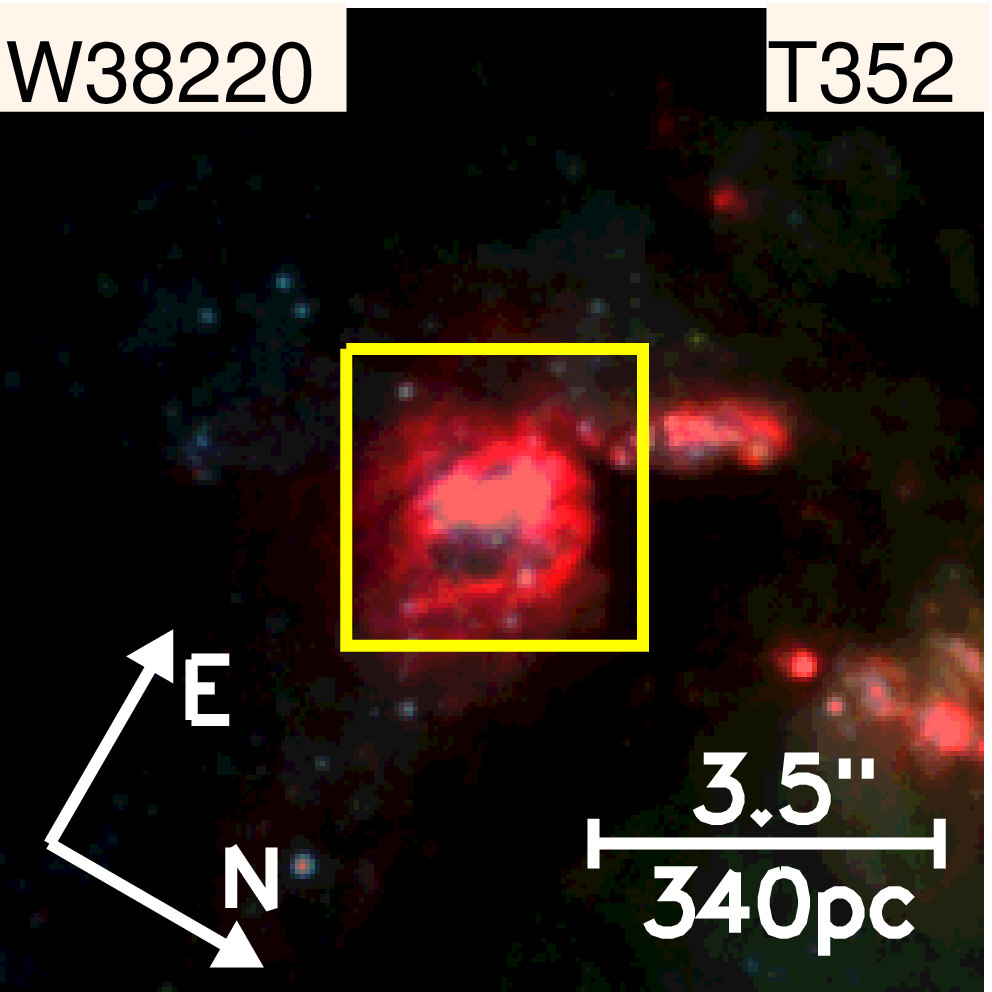}
\includegraphics[width=6cm]{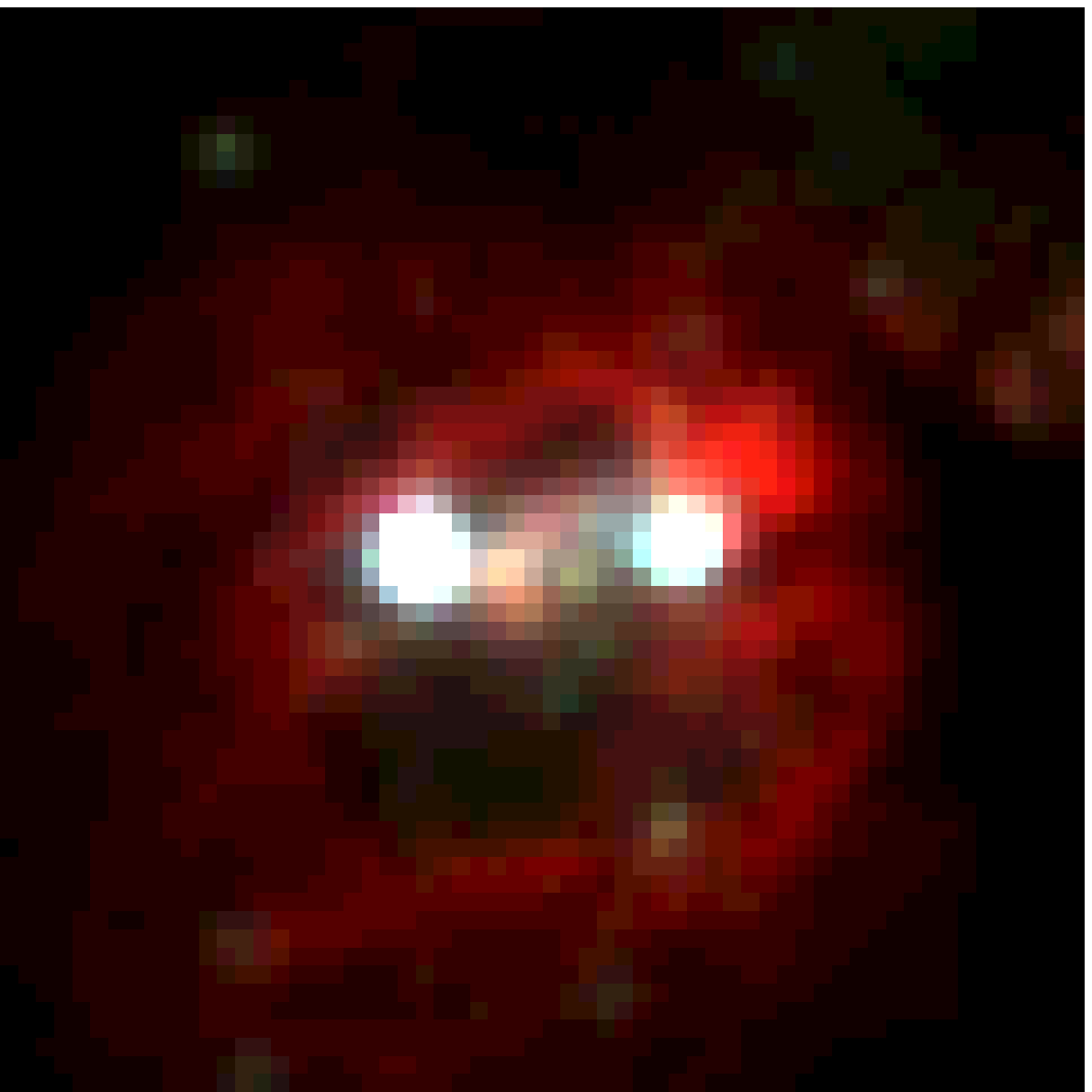}

\caption{{\bf Top panel:} A three colour HST/ACS images of T352 in the Antennae galaxies, blue, green and red represent images in the F435W, F550M, and F658N (H$\alpha$) filters, respectively.  The ID from Whitmore et al.~(2010) is also shown. {\bf Bottom panel:}  A zoom in on the region in the box in the top panel. }
\label{fig:t352}
\end{figure}

The Antennae galaxies, at a distance of $\sim20$~Mpc, host one of the largest and most well studied YMC populations in the local universe.  Optical and near-IR HST-based studies have found thousands of clusters, some with estimated masses near or significantly above $10^6$~\msun\ (e.g., Whitmore et al.~2010).  Ground based spectroscopic studies of some of the brightest and most massive clusters in the galaxy have largely confirmed age estimates based on broad-band photometry, and have identified extremely young YMCs ($<10$~Myr) through Wolf-Rayet star features in the integrated optical spectra (e.g., Bastian et al.~2009).  Based on the strong emission lines in H{\sc ii} regions, Bastian et al.~(2009) estimated that the metallicity of the galaxy is approximately solar.% has been estimated (Bastian et al. 2009).
%É.general info.  Studied by Bastian et al. (2009) and Whitmore et al. (2010).

Two of the most luminous clusters in these galaxies (W38220 and W38639 - IDs from Whitmore et al.~2010) appear to be extremely young ($\sim5-6$~Myr) and have already caused a large ionised bubble to be driven into the ISM.  The two clusters are not easily resolved from the ground, hence were studied together, labelled T352 in Bastian et al.~(2009), although are clearly resolved in HST imaging, and can be studied separately (Whitmore et al.~2010).

In Fig.~\ref{fig:t352} we show a three colour composite image of T352, based on HST/ACS imaging\footnote{See Bastian et al.~(2009) and Whitmore et al.~(2010) for a more detailed description of the images used.}.  The cluster on the left is W38220 with a photometrically determined mass of $9.2\times10^5$~\msun, an age of $5-6$~Myr, and an extinction of \av$=0.24$, and the cluster on the right is W38639 with a mass of $3\times10^5$~\msun, an age of $\sim6$~Myr, and an extinction of \av=$0.0$.  A clear ionised bubble is seen around both clusters, indicating that each has already removed any left over gas within them, and have been quite efficient in removing the ISM around them to a distance of $100-200$~pc.

The integrated spectra of the combined clusters shows clear Wolf-Rayet features, confirming the youth of the clusters.  Additionally, both clusters have blue $J-H$ (F110W - F160W) colours (0.30-0.34), indicating that neither cluster is dominated by red supergiants in the near-infrared yet.  Comparing these colours to the prediction of the simple stellar population models presented in Gazak et al.~(2013) leads to an age estimate of $\sim6$~Myr, in good agreement with the estimated age from optical photometry and the spectroscopically observed Wolf-Rayet features. 

We have estimated the size of each cluster using the same method as presented in Bastian et al.~(2009).  In that work, the individual clusters were not studied in detail since they could not be spatially resolved.  For the more massive cluster, W38220, the focus of the present work, we find an effective radius of $2.4^{+1.5}_{-1.0}$~pc. 

%\fbox{size estimate, done like in Bastian et al.~2009}

Another massive cluster, Knot S, has been studied both photometrically and spectroscopically.  While the cluster's light profile extends to $>300$~pc (Whitmore et al.~1999; Bastian et al.~2013c; similar to the ``standard model" of GC formation presented in D'Ercole et al.~2008), a relatively common feature for young high mass clusters, the ``core" of the cluster (the inner $\sim10$~pc) has been found to have an age of $\sim$5~Myr based on UV and optical photometry (Whitmore et al.~2010), as well as UV spectroscopy (Whitmore et al.~1999). The $J-H$ (F110W-F160W) colour (0.65) is redder than the previously discussed two clusters, suggesting a slightly older age.  Based on comparison between the observed $J-H$ and that predicted by Gazak et al.~(2013), we estimate an age of $\sim7$~Myr, in good agreement with previous methods.   Mengel et al.~(2008) have measured a dynamical mass for Knot~S of $3^{+1.2}_{-1.0} \times 10^6$~\msun, in good agreement with photometric based estimates ($1.6\times10^6$~\msun - Whitmore et al.~2010)\footnote{Throughout this paper, when errors associated with photometric mass determination were not given in the literature, we conservatively adopt errors of 50\%, which include uncertainties in the SSP models, extinction estimates, and distance uncertainties (e.g., Anders et al.~2004; Bastian et al.~2005).}.

\begin{figure}
\centering
\includegraphics[width=6cm]{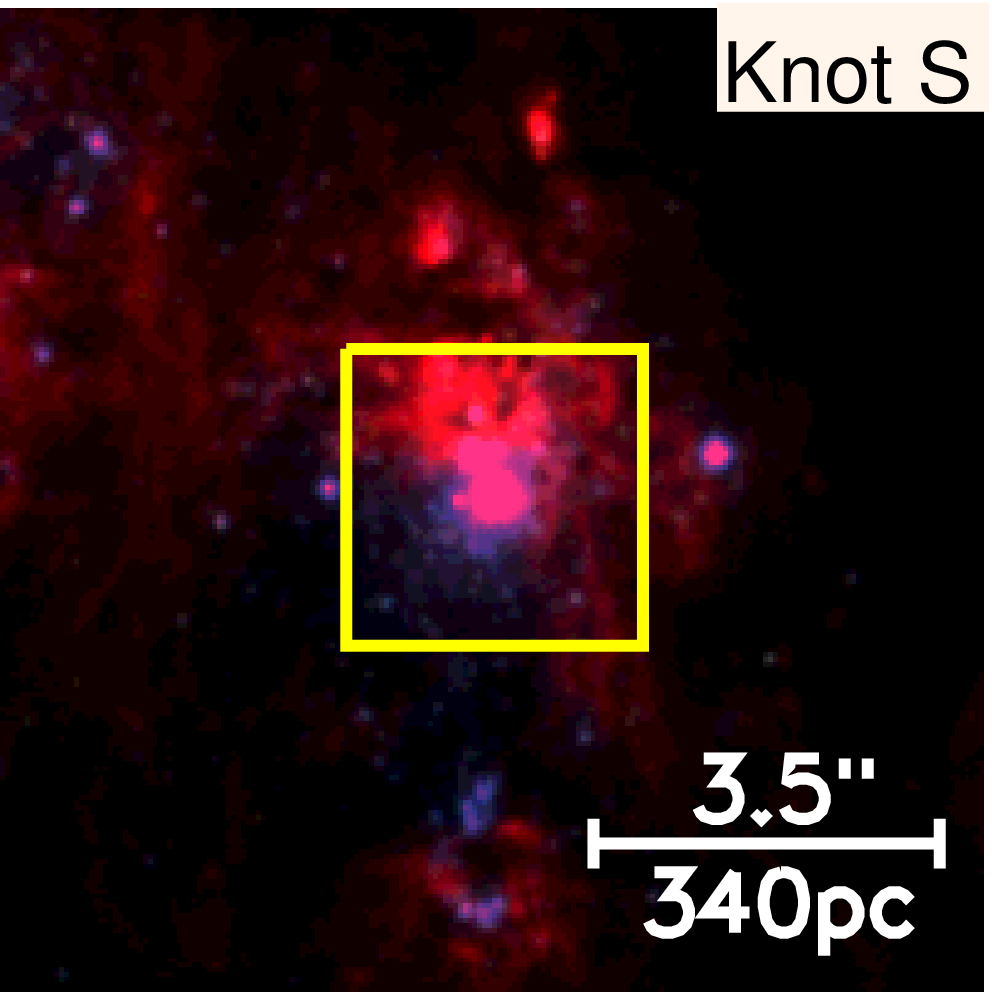}
\includegraphics[width=6cm]{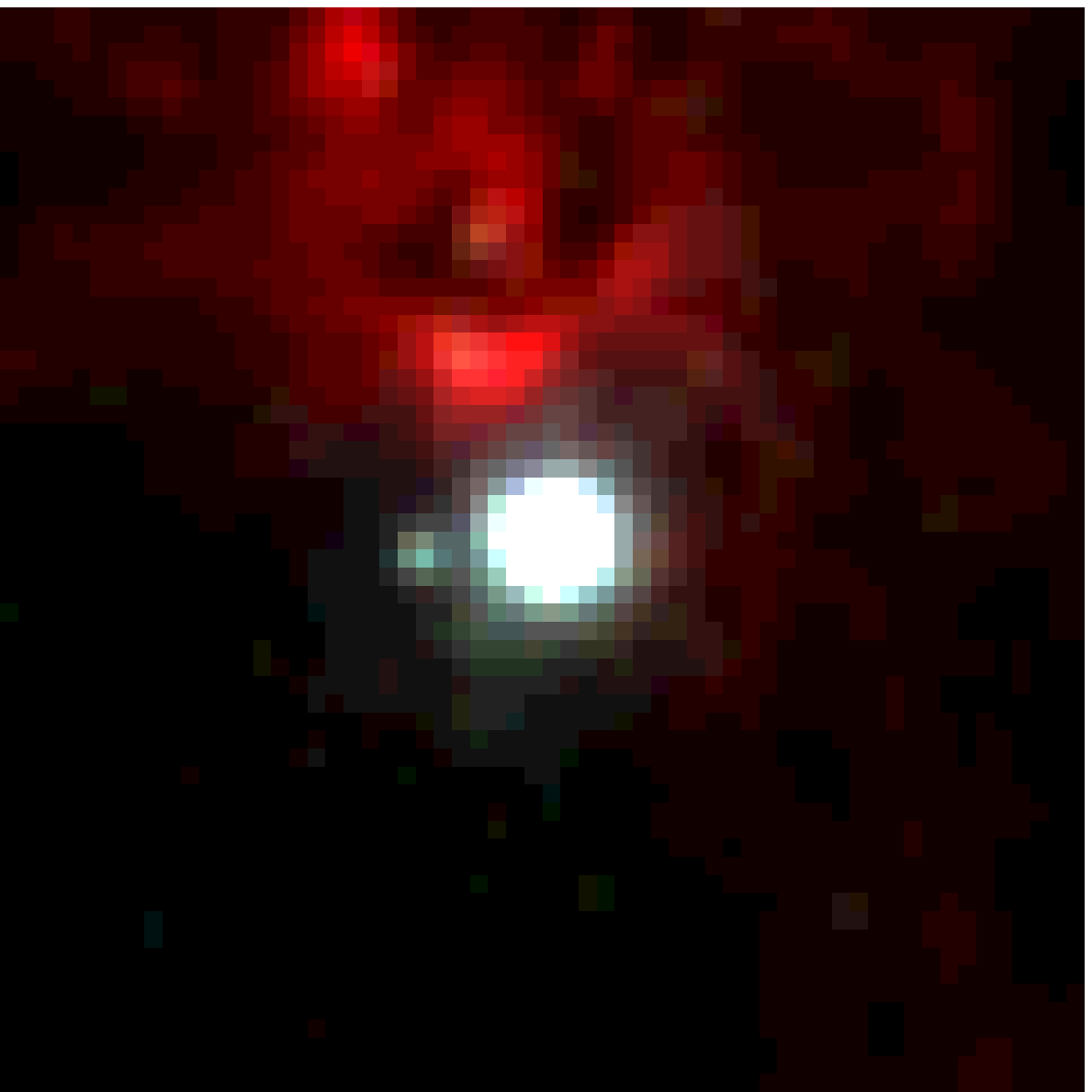}

\caption{The same as Fig.~\ref{fig:t352} but now showing Knot~S.  The orientation and scale are the same as Fig.~\ref{fig:t352}.}
\label{fig:knots}
\end{figure}

The extinction to Knot~S is low ($A_V < 0.4$ - Whitmore et al.~2010; Mengel et al. 2008).   Figure~\ref{fig:knots} shows a three colour composite image of Knot~S.  As seen in T352, the cluster has cleared out the surrounding ISM and it has become exposed, even at this high mass and young age, suggesting that the embedded phase for massive clusters lasts for less than 5~Myr. Whitmore et al.~(1999) use the H$\alpha$ bubble seen to the west of the cluster (with an estimated  size of 2.2$\pm0.2$"), along with their derived age, to derive an expansion speed of the bubble into the ISM, estimating it to be $\sim30$~km/s under the assumption that the bubble began expanding at the time of cluster formation (i.e., that the embedded phase is very short).  The estimated expansion velocity of the ionised bubbles is similar to that observed in other, slightly lower mass ($0.4-2\times10^6$~\msun), regions in the Antennae ($\sim40$~km/s - Bastian et al.~2006).  This lends support to the Whitmore et al. calculations and the assumptions that they adopted.  Hence, even very massive clusters are able to clear the gas within them, and drive large ionised bubbles into the ISM, soon after their formation (within 1-$\sim3$~Myr).

A summary of the properties of these two (and other) clusters is given in Table~\ref{tab:clusters}.
%Based on the strong emission lines in this region (i.e. the surrounding H{\sc ii} region), a metallicity of approximately solar has been estimated (Bastian et al. 2009).

\subsection{NGC~3256}

Another relatively nearby ongoing galaxy merger, NGC~3256, also hosts a large young cluster population (e.g., Zepf et al.~1999).  The population has been studied in detail using HST optical and UV imaging that found clusters with ages between a few Myr and a few hundred Myr (Goddard et al.~2010).  A handful of clusters have also been studied spectroscopically by Trancho et al.~(2007), who also determined a slightly super-solar metallicity ($\sim1.3~Z_{\odot}$) for a number of H{\sc ii} regions, using strong emission line methods.  One cluster, T2005, is particularly relevant to the current study, given its high mass ($\sim1.4 \times10^6$~\msun), youth ($<7$~Myr) and low extinction, $A_V \sim 0$~mag (Trancho et al.~2007).  The cluster displays strong Wolf-Rayet emission features, confirming its young age, and based on the emission line strengths, relative to other clusters in the galaxy, an age of $<5$~Myr can be estimated.

\begin{figure}
\centering
\includegraphics[width=6cm]{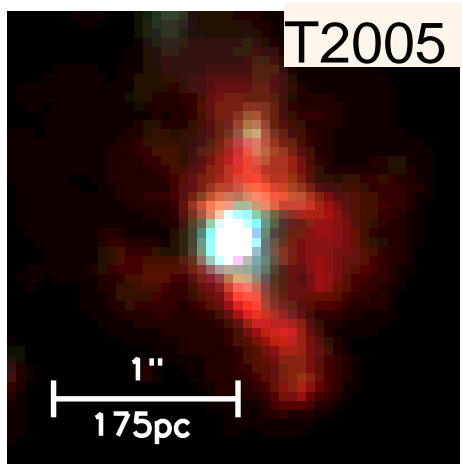}

\caption{Similar to Fig.~\ref{fig:t352}, but now showing cluster T2005 in NGC~3256. }
\label{fig:t2005}
\end{figure}

A three colour composite image of the cluster and the surrounding ISM is shown in Fig.~\ref{fig:t2005}.  The cluster appears to have cleared out the surrounding ISM, and is located within an ionised bubble with radius~$0.25"$, or $\sim40$~pc.  Like the previously discussed clusters, it appears that this young massive cluster has been capable of removing the ISM from within it on short timescales, and drive a large outflows into the ISM.

\section{Young High Mass Clusters in Low Metallicity Environments}
\label{sec:cluster23}

\subsection{ESO~338-IG04 - Cluster~23}

In order to most accurately compare massive clusters forming today with globular clusters, it is important to look at metal poor galaxies, with abundances in the same range of GCs.  In the nearby universe, Blue Compact Galaxies, with their low metallicities and high star-formation rates, offer a particularly insightful glimpse into GC formation (e.g., Adamo et al.~2011).  In this respect, a particularly interesting candidate for the current study is ESO~338-IG04 ({\"O}stlin et al.~2003; 2007; hereafter O3 and O7, respectively).  This low metallicity galaxy host a number of massive, young clusters, of which Cluster 23, with a mass of $5-16\times10^6$~\msun, an age of $6^{+4}_{-2}$~Myr, and an effective radius of $\sim5$~pc, is the most relevant to the present study.  The mass of the cluster has been determined photometrically (O3) and dynamically (O7) and overall the two methods find good agreement, however the dynamical mass of the cluster is systematically larger than that found from photometry, which is likely due to the contribution of binary stars to the measured velocity dispersion (Gieles et al.~2010).

The age of the cluster was found through matching the observed spectrum (in particular the wings of the Balmer lines, as the line centres contain significant amounts of emission) to simple stellar population models.  Additionally, matching individual stellar templates to the observed cluster spectrum resulted in a best fit temperature (and surface gravity) of O-type stars.  Furthermore, the strong He absorption lines suggest a young ($<10$~Myr) age.  The spectroscopic age agrees well with that estimated from HST based U, B, V,  and I photometry in comparison with SSP models (O3).

Due to the youth of this cluster, it is likely that it shares the metallicity of its host galaxy.  Based on emission lines, Bergvall \& {\"O}stlin~(2002) found an oxygen abundance of $(12 + log(O/H)=8.0)$,  or $\sim0.2$~\zsun, similar to that found based on a comparison of high resolution spectroscopy of Cluster 23 with predictions of SSP models (O7).  The extinction towards this cluster, as derived through integrated photometry, is very low, with E(B-V)$\sim0.05$.

%The youth of this cluster is confirmed by integrated photometry and high-resolution spectroscopy.  Based on the photometric estimates, which are confirmed through spectroscopy, the extinction towards this cluster is extremely low, with E(B-V)$\sim0.05$.  

\begin{figure}
\centering
\includegraphics[width=6cm]{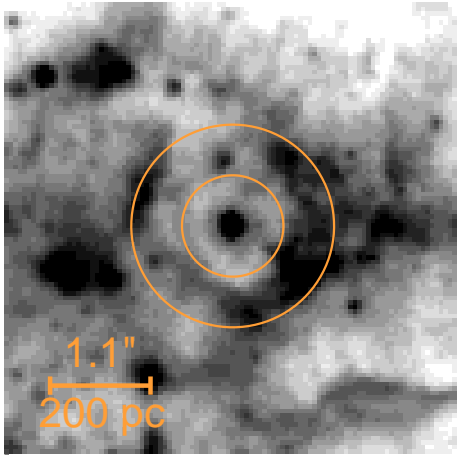}
\caption{An H$\alpha$ image (not continuum subtracted) of Cluster~23 in ESO~338-IG04.  The circles have radii of 100 and 200~pc.  Note the H$\alpha$ hole surrounding the cluster, with a ring of emission between 100 and 200~pc.}
\label{fig:cluster23}
\end{figure}

In Fig.~\ref{fig:cluster23} we show an HST H$\alpha$ image of Cluster 23 (not continuum subtracted).  A clear hole is seen around the cluster, surrounded by a ring of emission, a common morphology for young clusters that are expelling the gas within and around them through the combined effects of photoionisation, SNe, and stellar winds (e.g., Whitmore et al. 2011; Adamo et al.~2013; Hollyhead et al., in prep).  Superimposed are circles with radii of 100 and 200~pc (assuming a distance of 37.5~Mpc), showing that the ionised ring is currently located between 120 and 180~pc (in projection) from the cluster.  High resolution spectroscopy of the cluster has revealed a bubble with an expansion speed of $\sim40$~km/s (O7).  Assuming that this ionised expanding bubble is the same feature that we are observing, and also that the expansion speed has been constant, the bubble began expanding between 3 and 4.5~Myr ago.   

The age of the cluster is $6^{+4}_{-2}$~Myr, hence we conclude that this cluster has been able to rapidly clear any left-over gas from it's formation.  Comparing the estimated expansion age of the bubble ($3-4.5$~Myr) to the cluster age, we estimate that the cluster become exposed within $\sim1-7$~Myr of its formation.  A similar conclusion was reached by O7.

%(figure with the Halpha image, clear bubble) and thenÉ..{\"O}stlin et al.(2007) - ESO 338-IG04 - "Moreover, analysis of the [O iii] $\lambda5007$ and H$\alpha$ emission lines from the region near the younger cluster indicates that it is associated with a bubble expanding at ~40 km s-1. " 
%from nebular emission lines.  very similar to the metal rich GCs that do show 'multiple populations' in the stellar pops.

\subsection{Comparison with Milky Way Globular Clusters}
\label{sec:compare}

Bellini et al.~(2013) studied two globular clusters, NGC~6338 \& NGC~6441, associated with the bulge of the Galaxy, and found that both display evidence for multiple populations within them based on an analysis of the colour-magnitude diagrams.  These two clusters are relatively metal rich, with [Fe/H]$=-0.55$ and $-0.46$, respectively.  This shows that the phenomenon of multiple populations in GCs is not limited to metal poor clusters, indicating that the effect is independent of the environment at formation (i.e., in the bulge of a relatively massive galaxy like the Milky Way, or in small metal poor dwarfs that were later accreted) or their formation epoch (from redshifts of $\sim2$ to $5$ - cf. Renzini~2013).   

These two ancient GCs are more metal rich than  Cluster~1 in ESO~338-IG04.  Hence, in terms of its mass and metallicity, Cluster~1 can be considered a true proto-globular cluster, and theories for the formation and early evolution of GCs must satisfy the observational constraints provided by this, and other, YMCs.

%\section{Results}
%\label{sec:results}

\section{Discussion and Conclusions}
\label{sec:conclusions}

%Fast Rotating Massive Stars
In the scenario for globular cluster (GC) formation and early evolution put forward by Krause et al.~(2012; 2013), GCs remain embedded for the first $\sim30$~Myr of their lives.  Fast Rotating Massive Stars form enriched deccretion discs around them, which accrete primordial material that remains within the cluster (i.e., the left over material from the non-100\% star-formation efficiency).  A second generation of stars then forms within the deccretion discs, and any remaining primordial material is finally removed from the cluster after $\sim30$~Myr by the energy released by accretion onto dark remnants.  The authors argue that the mass of proto-GCs was too large for SNe to overcome the gravitational potential of the clusters, so such a scenario would only apply to very massive clusters.

We have tested this scenario by looking for young massive clusters (YMCs), with ages between a few and $\sim15$~Myr, that have already cleared out their natal material.  The Krause et al.~(2012; 2013) scenario predicts that no such clusters should be found.  By surveying the literature, we found that clusters with masses near or exceeding $10^6$~\msun\ become exposed (i.e., clear out their natal gas) within a few Myr after their formation, contradicting the Krause et al. predictions.  Many young clusters are surrounded by ionised bubbles, whose expansion speeds, as determined through high-resolution spectroscopy, and sizes, can be used to infer that the expansion began $1-3$~Myr after their formation.  Which mechanisms are the cause of the rapid removal of gas from young clusters is not entirely known.  As with lower mass clusters ($\lesssim10^5$~\msun, cf. Longmore et al.~2014), it appears that the gas is removed from the clusters before supernovae begin to explode ($7-10$~Myr for sub-solar metallicities, e.g., Heger et al.~2003).  Hence, ionisation and stellar winds from massive stars are the likely cause.  However, the observed bubble sizes (at young ages) appear to be larger than predicted by hydrodynamical models that include these types of stellar feedback (e.g., Dale et al.~2014).  This discrepancy may be due to differences in the assumed and actual initial conditions (i.e. a centrally concentrated rapid formation of a massive cluster, versus a more distributed stellar population).  Alternatively, it may be that extremely high mass stars ($>100$~\msun) may end their lives after $2-3$~Myr as pair-instability supernovae, adding a significant amount of energy into the surrounding ISM (e.g., Heger et al.~2003).  Additionally, the cause may be that the Dale et al.~(2014) simulations do not include radiation pressure, which may dominate in the dense regime of YMCs (e.g., Murray et al.~2010).  Future studies into the ionisation state (i.e., ionisation parameter mapping) of the ionised bubbles around young massive clusters may be able to differentiate between these scenarios (e.g., Pellegrini et al.~2012). 

One cluster in our sample is particularly useful in constraining the proposed models.  This cluster, ESO~338-IG04 Cluster~23, has an age of $6^{+4}_{-2}$~Myr, and a photometrically and dynamically determined mass of $5-16 \times10^6$~\msun.  The cluster is entirely exposed and sits in the centre of an ionised shell with radius $\sim150$~pc that is expanding at $\sim40$~km/s ({\"O}stlin et al.~2007).  Additionally, the metallicity of the ISM of the galaxy ($\sim0.2$~\zsun\ - {\"O}stlin et al.~2003) and the cluster itself ({\"O}stlin et al.~2007), is lower than some of the metal rich GCs in the Galaxy that are known to host multiple populations.  This suggests that metallicity differences between YMCs and GCs are not causing the two types of objects to follow divergent evolutionary paths.

The sample of clusters presented here is not meant to be complete nor exhaustive.  Instead, the sample is meant to show that many high mass clusters ($\gtrsim10^6$~\msun) with ages $<20$~Myr exist, that are no longer embedded within their natal gas cloud.  Based on these observations, it appears that high mass clusters begin expelling any gas left-over within them, and driving large scale ionisation bubbles into the surrounding ISM within $1-3$~Myr after their formation.  This is the similar to that observed in lower mass ($\lesssim10^5$~\msun) clusters (e.g., Longmore et al.~2014; Hollyhead et al.~in prep.), showing that the timescale for removing the gas from YMCs is largely independent of the cluster mass.  We conclude that observations of YMCs are in strong tension with the predictions of the early evolution of GCs in the Krause et al.~(2012; 2013) model.  However, we stress that the results presented here do not test whether FRMS are the source of the enriched material in GCs.  We have only tested the framework presented by Krause et al.~(2012; 2013) for how such material may be incorporated within a significant fraction of GC stars.

We note that the observations proposed here are consistent with both the AGB and early disc accretion (EDA) scenarios.  Both scenarios invoke the rapid expulsion ($\lesssim3$~Myr) of any gas left over from the formation of the initial population of stars.  In the AGB scenario, the young GCs need to accrete large amounts of primordial material from their surroundings, beginning $\sim30$~Myr after their formation, which was not tested in the current study.  In the EDA scenario, the remaining natal gas is removed within the first $1-3$~Myr of a cluster's life, and no further gas is accreted onto the cluster from the surroundings.  While some dense gas is expected to exist within the young GCs for short time, i.e., material from interacting binary stars or FRMSs, the vast majority of the cluster is gas free at any given time.  Hence, the EDA model does not predict that the young GCs would appear embedded at any time after they clear away their natal gas.

However, the observations presented here may also impact the AGB scenario, as the young clusters have been able to remove any pristine gas from within or near the cluster, out to distances $>200$~pc. It is likely that at these distances, the gas is no longer bound to the cluster, but rather will be dominated by the gravitational potential of the host galaxy.  Hence, it is not clear how the cluster will be able to re-accrete the large amounts of pristine material (in order to form the 2nd generation of stars), that is required in the AGB scenario.  This is especially difficult, considering the fine-tuning in the accretion time necessary to match the models to the observations, i.e. that the first stars of the ``2nd generation" form from entirely enriched/processed material, and that the pristine material is not brought into the cluster until $\sim50$~Myr after the first generation forms. (e.g., D'Ercole, D'Antona, \& Vesperini~2011).

%Of particular relevance is the cluster ESO~338-IG04 which has a metallicity below (at least) two of the Milky Way globular clusters that have multiple populations.

\section*{Acknowledgments}

We thank Martin Krause, Corinne Charbonnel and the anonymous referee for helpful discussions and comments on the draft.   NB is partially funded by a Royal Society University Research Fellowship. The results presented here are partially based on observations made with the NASA/ESA Hubble Space Telescope, and obtained from the Hubble Legacy Archive, which is a collaboration between the Space Telescope Science Institute (STScI/NASA), the Space Telescope European Coordinating Facility (ST-ECF/ESA) and the Canadian Astronomy Data Centre (CADC/NRC/CSA).

\bsp
\label{lastpage}

\begin{thebibliography}{99}

\bibitem[Adamo et al.(2011)]{2011MNRAS.417.1904A} Adamo, A., {\"O}stlin, 
G., \& Zackrisson, E.\ 2011, MNRAS, 417, 1904 

\bibitem[Adamo et al.(2012)]{2012MNRAS.426.1185A} Adamo, A., Smith, L.~J., 
Gallagher, J.~S., et al.\ 2012, MNRAS, 426, 1185 

\bibitem[Anders et al.(2004)]{2004MNRAS.347..196A} Anders, P., Bissantz, 
N., Fritze-v.~Alvensleben, U., \& de Grijs, R.\ 2004, MNRAS, 347, 196 


\bibitem[Bastian et 
al.(2005)]{2005A&A...431..905B} Bastian, N., Gieles, M., Lamers, H.~J.~G.~L.~M., Scheepmaker, R.~A., \& de Grijs, R.\ 2005, A\&A, 431, 905 


\bibitem[Bastian et 
al.(2006)]{2006A&A...445..471B} Bastian, N., Emsellem, E., Kissler-Patig, M., \& Maraston, C.\ 2006, A\&A, 445, 471 

\bibitem[Bastian(2008)]{2008MNRAS.390..759B} Bastian, N.\ 2008, MNRAS, 
390, 759 

\bibitem[Bastian et al.(2009)]{2009ApJ...701..607B} Bastian, N., Trancho, 
G., Konstantopoulos, I.~S., \& Miller, B.~W.\ 2009, ApJ, 701, 607 

\bibitem[Bastian et al.(2013)]{2013MNRAS.436.2852B} Bastian, N., 
Cabrera-Ziri, I., Davies, B., \& Larsen, S.~S.\ 2013a, MNRAS, 436, 2852 (Paper I)

\bibitem[Bastian et al.(2013)]{2013MNRAS.436.2398B} Bastian, N., Lamers, 
H.~J.~G.~L.~M., de Mink, S.~E., et al.\ 2013b, MNRAS, 436, 2398 

\bibitem[Bastian et al.(2013)]{2013MNRAS.431.1252B} Bastian, N., Schweizer, 
F., Goudfrooij, P., Larsen, S.~S., 
\& Kissler-Patig, M.\ 2013c, MNRAS, 431, 1252 


\bibitem[Bastian \& Strader(2014)]{bastian14} Bastian, N. \& Strader, J.~2014, MNRAS, in press (arXiv:1407.2726) (Paper III)


\bibitem[Bellini et al.(2013)]{2013ApJ...765...32B} Bellini, A., Piotto, 
G., Milone, A.~P., et al.\ 2013, ApJ, 765, 32 

\bibitem[Bergvall \& {\"O}stlin(2002)]{2002A&A...390..891B} Bergvall, N., {\"O}stlin , G.\ 2002, A\&A, 390, 891 

\bibitem[Brodie 
\& Strader(2006)]{2006ARA&A..44..193B} Brodie, J.~P., \& Strader, J.\ 2006, ARAA, 44, 193 


\bibitem[Cabrera-Ziri et al.(2014)]{2014MNRAS.441.2754C} Cabrera-Ziri, I., 
Bastian, N., Davies, B., et al.\ 2014, MNRAS, 441, 2754 (Paper II)

\bibitem[Conroy \& Spergel(2011)]{2011ApJ...726...36C} Conroy, C., \& Spergel, D.~N.\ 2011, ApJ, 726, 36 

\bibitem[Dale et al.(2014)]{2014MNRAS.442..694D} Dale, J.~E., Ngoumou, J., 
Ercolano, B., \& Bonnell, I.~A.\ 2014, MNRAS, 442, 694 


\bibitem[D'Antona et al.(2013)]{2013MNRAS.434.1138D} D'Antona, F., Caloi, 
V., D'Ercole, A., et al.\ 2013, MNRAS, 434, 1138 


\bibitem[Decressin et 
al.(2007)]{2007A&A...464.1029D} Decressin, T., Meynet, G., Charbonnel, C., Prantzos, N., \& Ekstr{\"o}m, S.\ 2007a, A\&A, 464, 1029 


\bibitem[Decressin et 
al.(2007)]{2007A&A...475..859D} Decressin, T., Charbonnel, C., \& Meynet, G.\ 2007b, A\&A, 475, 859 

\bibitem[D'Ercole et al.(2008)]{2008MNRAS.391..825D} D'Ercole, A., 
Vesperini, E., D'Antona, F., McMillan, S.~L.~W., 
\& Recchi, S.\ 2008, MNRAS, 391, 825 

\bibitem[D'Ercole et al.(2011)]{2011MNRAS.415.1304D} D'Ercole, A., 
D'Antona, F., \& Vesperini, E.\ 2011, MNRAS, 415, 1304 


\bibitem[Elmegreen et al.(2000)]{2000ApJ...535..748E} Elmegreen, B.~G., 
Efremov, Y.~N., \& Larsen, S.\ 2000, ApJ, 535, 748 

\bibitem[Gazak et al.(2013)]{2013MNRAS.430L..35G} Gazak, J.~Z., Bastian, N., Kudritzki, R.-P., et al.\ 2013, MNRAS, 430, L35 

\bibitem[Gazak et al.(2014)]{2014ApJ...787..142G} Gazak, J.~Z., Davies, B., 
Bastian, N., et al.\ 2014, ApJ, 787, 142 

\bibitem[Gieles et al.(2010)]{2010MNRAS.402.1750G} Gieles, M., Sana, H., 
\& Portegies Zwart, S.~F.\ 2010, MNRAS, 402, 1750 

\bibitem[Goddard et al.(2010)]{2010MNRAS.405..857G} Goddard, Q.~E., 
Bastian, N., \& Kennicutt, R.~C.\ 2010, MNRAS, 405, 857 

\bibitem[Goudfrooij et al.(2011)]{2011ApJ...737....4G} Goudfrooij, P., 
Puzia, T.~H., Chandar, R., \& Kozhurina-Platais, V.\ 2011, ApJ, 737, 4 

\bibitem[Heckman 
\& Leitherer(1997)]{1997AJ....114...69H} Heckman, T.~M., \& Leitherer, C.\ 1997, AJ, 114, 69 


\bibitem[Heger et al.(2003)]{2003ApJ...591..288H} Heger, A., Fryer, C.~L., 
Woosley, S.~E., Langer, N., \& Hartmann, D.~H.\ 2003, ApJ, 591, 288 


\bibitem[Hunter et al.(2000)]{2000AJ....120.2383H} Hunter, D.~A., 
O'Connell, R.~W., Gallagher, J.~S., 
\& Smecker-Hane, T.~A.\ 2000, AJ, 120, 2383 

\bibitem[Kobulnicky 
\& Skillman(1997)]{1997ApJ...489..636K} Kobulnicky, H.~A., \& Skillman, E.~D.\ 1997, ApJ, 489, 636 

\bibitem[Krause et 
al.(2012)]{2012A&A...546L...5K} Krause, M., Charbonnel, C., Decressin, T., et al.\ 2012, A\&A, 546, L5 

\bibitem[Krause et 
al.(2013)]{2013A&A...552A.121K} Krause, M., Charbonnel, C., Decressin, T., Meynet, G., \& Prantzos, N.\ 2013, A\&A, 552, A121 

\bibitem[Kruijssen(2012)]{2012MNRAS.426.3008K} Kruijssen, J.~M.~D.\ 2012, 
MNRAS, 426, 3008 


\bibitem[Kruijssen(2014)]{2014arXiv1407.2953K} Kruijssen, J.~M.~D.\ 2014, Classical and Quantum Gravity (in press - arXiv:1407.2953)


%\bibitem[Larsen et al.(2001)]{2001ApJ...556..801L} Larsen, S.~S., Brodie, 
%J.~P., Elmegreen, B.~G., et al.\ 2001, ApJ, 556, 801 

\bibitem[Larsen et al.(2001)]{2001ApJ...556..801L} Larsen, S.~S., Brodie, 
J.~P., Elmegreen, B.~G., et al.\ 2001, ApJ, 556, 801 

\bibitem[Larsen et al.(2004)]{2004AJ....128.2295L} Larsen, S.~S., Brodie, 
J.~P., \& Hunter, D.~A.\ 2004, AJ, 128, 2295 

\bibitem[Larsen et al.(2006)]{2006AJ....131.2362L} Larsen, S.~S., Brodie, 
J.~P., \& Hunter, D.~A.\ 2006a, AJ, 131, 2362 

\bibitem[Larsen et al.(2006)]{2006MNRAS.368L..10L} Larsen, S.~S., Origlia, 
L., Brodie, J.~P., \& Gallagher, J.~S.\ 2006b, MNRAS, 368, L10 

\bibitem[Larsen et al.(2008)]{2008MNRAS.383..263L} Larsen, S.~S., Origlia, 
L., Brodie, J., \& Gallagher, J.~S.\ 2008, MNRAS, 383, 263 

\bibitem[Larsen et 
al.(2011)]{2011A&A...532A.147L} Larsen, S.~S., de Mink, S.~E., Eldridge, J.~J., et al.\ 2011, A\&A, 532, A147 

\bibitem[Larsen et 
al.(2012)]{2012A&A...546A..53L} Larsen, S.~S., Brodie, J.~P., \& Strader, J.\ 2012, A\&A, 546, A53 

\bibitem[Larsen et 
al.(2014)]{2014A&A...565A..98L} Larsen, S.~S., Brodie, J.~P., Forbes, D.~A., \& Strader, J.\ 2014, A\&A, 565, A98 

\bibitem[Lee 
\& Skillman(2004)]{2004ApJ...614..698L} Lee, H., \& Skillman, E.~D.\ 2004, ApJ, 614, 698 

\bibitem[Longmore et al.(2014)]{2014arXiv1401.4175L} Longmore, S.~N., 
Kruijssen, J.~M.~D., Bastian, N., et al.\ 2014, PPVI, in press (arXiv:1401.4175) 

\bibitem[Maoz et al.(2001)]{2001ApJ...554L.139M} Maoz, D., Ho, L.~C., 
\& Sternberg, A.\ 2001, ApJL, 554, L139 

\bibitem[Mengel et 
al.(2008)]{2008A&A...489.1091M} Mengel, S., Lehnert, M.~D., Thatte, N.~A., et al.\ 2008, A\&A, 489, 1091 

\bibitem[Meurer et al.(1992)]{1992AJ....103...60M} Meurer, G.~R., Freeman, 
K.~C., Dopita, M.~A., \& Cacciari, C.\ 1992, AJ, 103, 60 


\bibitem[Moll et al.(2007)]{2007MNRAS.382.1877M} Moll, S.~L., Mengel, S., 
de Grijs, R., Smith, L.~J., \& Crowther, P.~A.\ 2007, MNRAS, 382, 1877 

\bibitem[Murray et al.(2010)]{2010ApJ...709..191M} Murray, N., Quataert, 
E., \& Thompson, T.~A.\ 2010, ApJ, 709, 191 

\bibitem[{\"O}stlin et 
al.(2003)]{2003A&A...408..887O} {\"O}stlin, G., Zackrisson, E., Bergvall, N., R{\"o}nnback, J.\ 2003, A\&A, 408, 887 

\bibitem[{\"O}stlin et 
al.(2007)]{2007A&A...461..471O} {\"O}stlin, G., Cumming, R.~J., \& Bergvall, N.\ 2007, A\&A, 461, 471 

\bibitem[Pellegrini et al.(2012)]{2012ApJ...755...40P} Pellegrini, E.~W., 
Oey, M.~S., Winkler, P.~F., et al.\ 2012, ApJ, 755, 40 


\bibitem[Prantzos 
\& Charbonnel(2006)]{2006A&A...458..135P} Prantzos, N., \& Charbonnel, C.\ 2006, A\&A, 458, 135 

\bibitem[Renzini(2013)]{2013MmSAI..84..162R} Renzini, A.\ 2013, Memorie della Societa Astronomica Italiana, 
84, 162 

\bibitem[Seale et al.(2012)]{2012ApJ...751...42S} Seale, J.~P., Looney, 
L.~W., Wong, T., et al.\ 2012, ApJ, 751, 42 

\bibitem[Smith 
\& Gallagher(2001)]{2001MNRAS.326.1027S} Smith, L.~J., \& Gallagher, J.~S.\ 2001, MNRAS, 326, 1027 


\bibitem[Sollima et al.(2013)]{2013MNRAS.433.1276S} Sollima, A., Gratton, 
R.~G., Carretta, E., Bragaglia, A., 
\& Lucatello, S.\ 2013, MNRAS, 433, 1276 

\bibitem[Trancho et al.(2007)]{2007ApJ...664..284T} Trancho, G., Bastian, 
N., Miller, B.~W., \& Schweizer, F.\ 2007, ApJ, 664, 284 

\bibitem[Whitmore et al.(1999)]{1999AJ....118.1551W} Whitmore, B.~C., 
Zhang, Q., Leitherer, C., et al.\ 1999, AJ, 118, 1551 

\bibitem[Whitmore et al.(2010)]{2010AJ....140...75W} Whitmore, B.~C., 
Chandar, R., Schweizer, F., et al.\ 2010, AJ, 140, 75 

\bibitem[Whitmore et al.(2011)]{2011ApJ...729...78W} Whitmore, B.~C., 
Chandar, R., Kim, H., et al.\ 2011, ApJ, 729, 78 

\bibitem[Zepf et al.(1999)]{1999AJ....118..752Z} Zepf, S.~E., Ashman, 
K.~M., English, J., Freeman, K.~C., \& Sharples, R.~M.\ 1999, AJ, 118, 752 

\end{thebibliography}
\end{document}